\documentclass[aps,pre,twocolumn,amsmath,superscriptaddress,longbibliography]{revtex4-2}

\newcommand{\bea}{\begin{eqnarray}}
\newcommand{\eea}{\end{eqnarray}}
\newcommand{\beq}{\begin{equation}}
\newcommand{\eeq}{\end{equation}}

\usepackage{amsmath}
\usepackage[urlcolor=blue,colorlinks=true,citecolor=blue,linkcolor=blue,pdfstartview={FitH},bookmarks=false]{hyperref}
\usepackage{appendix}
\usepackage{graphicx}
\usepackage{longtable}
\usepackage{epsfig}
\usepackage{dcolumn}
\usepackage{bm}
\usepackage{bbm}
\usepackage{amssymb}
\usepackage{multirow}
\usepackage{times,color}
\usepackage{hyperref}
\usepackage{amsmath}
\usepackage{color}
\usepackage{subfigure}
\usepackage{overpic}
\usepackage{float}
\usepackage{epstopdf}
\usepackage{braket} 
\usepackage{diagbox}
\usepackage{color}
\usepackage{url}
\usepackage{xcolor}
\begin{document}
\title{Quantum Criticality of Generalized Aubry-Andr\'{e} Models with Exact Mobility Edges Using Fidelity Susceptibility}

\author{Yu-Bin Liu}
\thanks{These authors contributed equally to this work.}
\affiliation{College of Physics, Nanjing University of Aeronautics and Astronautics, Nanjing, 211106, China}
\affiliation{Key Laboratory of Aerospace Information Materials and Physics (NUAA), MIIT, Nanjing 211106, China}
 \author{Wen-Yi Zhang}
 \thanks{These authors contributed equally to this work.}
\affiliation{College of Physics, Nanjing University of Aeronautics and Astronautics, Nanjing, 211106, China}
\affiliation{Key Laboratory of Aerospace Information Materials and Physics (NUAA), MIIT, Nanjing 211106, China}
\author{Tian-Cheng Yi}
\affiliation{Department of physics, Zhejiang Sci-Tech University, Hangzhou 310018, China}
\author{Liangsheng Li}
\affiliation{National Key Laboratory of Scattering and Radiation, Beijing 100854, China}

\author{Maoxin Liu}
 \affiliation{School of Systems Science  $\&$ Institute of Nonequilibrium Systems, Beijing Normal University, Beijing 100875, China}
             
\author{Wen-Long You}
\email{wlyou@nuaa.edu.cn}
\affiliation{College of Physics, Nanjing University of Aeronautics and Astronautics, Nanjing, 211106, China}
\affiliation{Key Laboratory of Aerospace Information Materials and Physics (NUAA), MIIT, Nanjing 211106, China}

\begin{abstract}
In this study, we explore the quantum critical phenomena in 
generalized  Aubry-Andr\'{e} models, with a particular focus on the scaling behavior at various filling states. Our approach involves using quantum fidelity susceptibility to precisely identify the mobility edges in these systems.  Through a finite-size scaling analysis of the fidelity susceptibility, we are able to determine both the correlation-length critical exponent and the dynamical critical exponent at the critical point of the generalized Aubry-Andr\'{e} model. Based on the Diophantine equation conjecture, we can determines the number of subsequences of the Fibonacci sequence and the corresponding  scaling functions for a specific filling fraction, as well as the universality class.  Our findings demonstrate the effectiveness of employing the generalized fidelity susceptibility for the analysis of  unconventional quantum criticality and the associated universal information of quasiperiodic systems in cutting-edge  quantum simulation experiments. 
\end{abstract}

\maketitle

\section{Introduction}
\label{intro}

In past decades, the study of phase transitions in disordered systems has gained upsurging interest in condensed matter physics. Anderson's pioneering theory of localization in noninteracting disordered systems has achieved remarkable success~\cite{Anderson1958}. Such systems may exhibit many-body  localization (MBL) in the presence of many-body interactions~\cite{Anderson_MBL_1980}. Recently, the phenomenon of MBL was investigated extensively from both theoretical and experimental perspectives~\cite{Basko2006,Gornyi2005,Abanin2019}. A key insight of current phenomenological theory is that MBL systems exist an emergent integrability, which is characterized by  a set of quasilocal integrals of motion. The emergent integrability plays a similar role of exact integrability in integral systems, and leads to a breakdown of thermalization in isolated systems. Interestingly, the strong ergodicity breaking arising from integrability and disorder 
fails to describe the recent discovery of quantum many-body scars, which triggered intensive investigations of weak ergodicity breaking in a variety of quantum systems ~\cite{You2022,zhang2023quantum,Serbyn2021}.  Compelling theoretical and numerical evidence  suggests the phenomenon of localization and non-ergodicity can exist in quasiperiodic models.  Compared to random disorder, which poses challenges in experimental engineering~\cite{Nieuwenburg2019}, incommensurate potential systems are more accessible for  experimental platforms, particularly in the context of ultra-cold atoms~\cite{Michael2015}, photonic crystals~\cite{Negro_photonic_2003} and cavity polaritons~\cite{DeGottardi2013,Ni2019},  providing a fertile ground for investigating incommensurate potential driven localization.  
The dispensation of disorder averaging has conduced to an in-depth theoretical investigations  in quasiperiodic systems.
Despite many physical features in common, such as multifractal wave functions and exponentially-diverging localization length, the critical properties of the localization transition in random and incommensurate models are rather different. As the random matrix theory is not applicable to describing  incommensurate models~\cite{Brody1981}, there are no rigorous bounds on the critical exponents 
concerning the localization transition~\cite{Chayes1986}.  Considering all eigenstates of Anderson model will be localized for an infinitesimal disorder strength in one and two dimensions (1D, 2D), a distinct feature of incommensurate model is the access to a localization-delocalization transition in 1D. Therefore,  quasiperiodic models open the door to effectively simulate some properties of Anderson localization that occur generically in three dimensions (3D) and offer a fascinating platform to explore unconventional phenomena, including fractal bands~\cite{Jia-Qi2021, Tanese2014}, mobility edge (ME)~\cite{Bloch_ME_2018}, Bose glass~\cite{Yao2020}, topological phases~\cite{Zhang2022}, and non-Hermitian Floquet quasicrystals~\cite{Weidemann2022}. 
\par
 Among a diversity of quasiperiodic models, the Aubry-Andr\'{e} (AA) model and its generalizations are drawing increasing attention nowadays due to its experimental realizations~\cite{Vardeny2013,Li2023}.  The 1D AA model was originally derived from the reduction of a square-lattice Hofstadter Hamiltonian describing 
conduction electrons in the presence of a uniform magnetic field~\cite{Harper1955}. 
A dual transformation between the coordinate and momentum spaces will yield the same Hamiltonian with hopping and potential amplitudes interchanged. As a consequence of the inherent self-duality, all the eigenstates of the system undergo Anderson localization transition from an extended phase (EP) to a localized phase (LP) at a critical value of the incommensurate potential strength~\cite{Wei2019,Sinha2019}.  
At the transition, the energy spectrum  showcases the renowned Hofstadter butterfly pattern~\cite{Hofstadter1976}. The multifractal
gap structure that ultimately evolves into the localization of the particles above the transition point.
During the past few years, inspiring progress is made by tailoring the incommensurate potential to introduce a single-particle ME into the AA model~\cite{Biddle2010}, which separates localized and delocalized eigenstates  at a critical energy.
Typical variants encompass slowly varying Aubry-Andr\'{e} model~\cite{Sarma1988,Sarma1990}, mosaic Aubry-Andr\'{e} model~\cite{Wang2020} and generalized Aubry-Andr\'{e} model (GAA)~\cite{Ganeshan2015}.

 Motivated by the recent experimental progress, specifically the successful simulation of the GAA model by using quantum Cs  gas~\cite{Wang2022} and using synthetic lattices of laser-coupled atomic momentum modes~\cite{Fangzhao2021}, 
we consider critical properties of the GAA model that has a single-particle ME at a finite filling. Our primary focus is on the scaling behavior of the fidelity susceptibility across the localization transition between the EP and the EP at different filling fractions.
{In the context of the Widom scaling hypothesis~\cite{Widom1965Equation}, which is foundational to the field of critical phenomena, it is posited that a complete characterization of a universality class of a phase transition can be achieved using just two independent critical exponents.} The finite-size scaling of the generalized participation indicated that the correlation-length critical exponent $\nu$ = 1 for the AA model, in which an incommensurate on-site potential is characterized by an irrational number $\alpha$.
 A second independent critical exponent thus plays a decisive role in determining the universality class. 
The ground state universality class of quasiperiodic models has been thoroughly investigated using a renormalization group scheme~\cite{Szabo2018,Thakurathi2012}.   It has been recognized that the universality class of the
ground state depended solely on the continued-fraction
expansion of  $\alpha$~\cite{Hashimoto_1992}.  Furthermore, the dynamic critical exponent $z$ exhibits a significant dependence on irrational numbers.   The value of $z$ at different filling fractions dictate the scaling behavior of the corresponding energy spectrum~\cite{Evangelou2000,Hiramoto1989,Cestari2011}, which proves valuable in comprehending both the multifractal characteristics of the system~\cite{Tang1986} and the properties associated with multiparticle ground-state transitions.
 Notably, the relationship between $z$, $\alpha$, and the filling fractions $\rho$ is less thoroughly understood. 
A strategy using  generalized fidelity susceptibility (GFS) to obtain the dynamical exponent $z$ has been developed for the lowest state in the both CP-LP~\cite{Lv2022_pra} and EP-CP transitions~\cite{Lv2022_prb}  of the $p$-wave-paired 
AA model.  

\par
This paper is organized as follows. In Sec. \ref{sec:MODEL},  we briefly present the GAA models and the phase diagram. In Sec. \ref{sec:QPTGAA}, we show the ME of GAA model by inverse participation ratio and fidelity susceptibility. In Sec. \ref{sec:Scaling hypothesis}, we review the scaling hypothesis of  fidelity susceptibility and present the numerical results of the critical behaviour in the GAA model. We also unravel the relationship between the Diophantine equation and the filling fractions.  The summary and conclusion are
given in Sec. \ref{Summary and conclusions}.

\section{MODEL HAMILTONIAN}
\label{sec:MODEL}
\begin{figure}[t]
\centering
\includegraphics[width=\columnwidth]{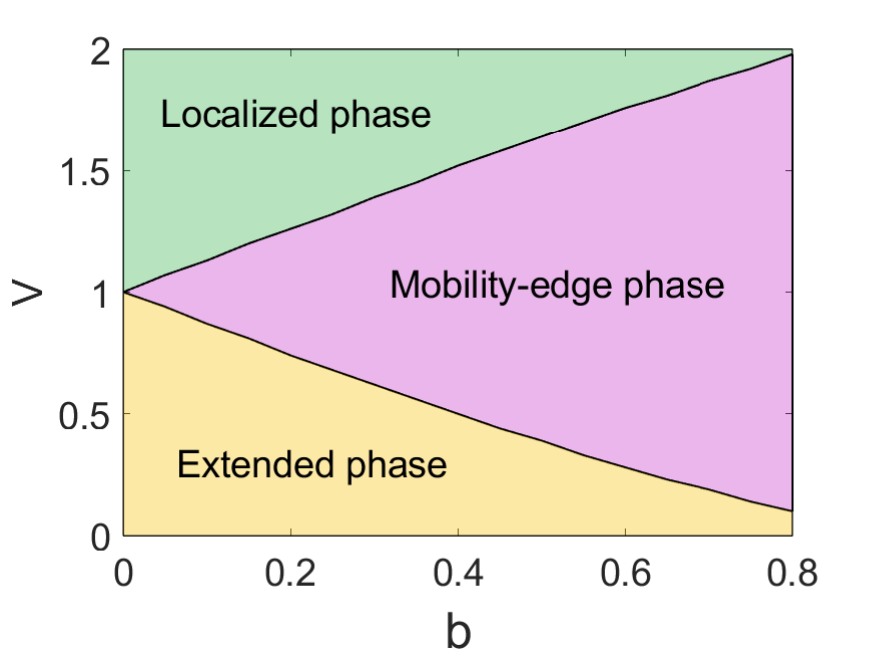}
 \caption{Phase diagram of the GAA model as a function of the incommensurate potential strength $V$ and the deformation parameter $b$. 
 The  green, yellow and pink regions represent the completely localized, completely extended  and mobility-edge phases, respectively.
 In the mobility-edge phase,  the extended states and localized states coexist.  Here we choose $N=1597$.
 }
\label{phase_diagram_GAA}
\end{figure}

We consider a family of GAA models described by the following Hamiltonian
\begin{eqnarray}
\hat{H}= -t\sum_{j=1}^{N}(\hat{c}_{j}^{\dagger}\hat{c}_{j+1}+\hat{c}_{j+1}^{\dagger}\hat{c}_{j})+\sum_{j=1}^{N}V_j\hat{c}_j^{\dagger}\hat{c}_{j},
\label{equ:Hamiltonian}
\end{eqnarray}
in which the operator $c_j^\dagger$ ($c_j$) represents the creation (annihilation) operator at site $j$ among total $N$ lattice sites, and the hopping amplitude $t$ is set to be the unit energy scale,  i.e., $t=1$.  
The on-site potential $V_j$ is characterized by a modulation strength $V$, a wave number  $\alpha$, a deformation parameter $b$ and a random phase $\phi$,  given by  
\begin{eqnarray}
V_j=2V\frac{\cos(2\pi\alpha j+\phi )}{1-b\cos(2\pi\alpha j +\phi)}.
\end{eqnarray}
When $\alpha$ takes an irrational 
value, the modulation becomes quasi-periodic. In this work, we employ $\alpha=(\sqrt{5}-1)/{2}$. 
The on-site potential is a continuous function of the parameter $b$, where $b\in[0,1)$. When $b=0$, the model (\ref{equ:Hamiltonian})
reduces to the standard AA model. For $V<t$ all the single-particle eigenstates are extended, while they become localized simultaneity for
$V>t$. At $V=t$,  the system undergoes the Anderson localization transition, leading to all eigenstates becoming critical and exhibiting multifractal behavior. This localization transition at $V=t$ is a consequence of a self-dual symmetry present in the system, which indicates the absence of the ME within the system. For $b\neq 0$, the deformed AA 
Hamiltonian in Eq. (\ref{equ:Hamiltonian}) respects an energy-dependent
duality symmetry, which ensures the existence of localization transition. 
The GAA duality warrants the occurrence of the ME described by a closed-form analytical expression:
\begin{eqnarray}
bE=2(t-V),
\label{equ:mathsolution}
\end{eqnarray}
 where $E$ represents the single-particle energy. 
 {According to Eq. (\ref{equ:mathsolution}), we depict the phase diagram} in Fig. \ref{phase_diagram_GAA}, 
in which the single-particle eigenstates are categorized into three types: completely localized, completely extended and mobility-edge phases.
In particular,  low-energy states are extended in the mobility-edge phase. In this study, 
{we investigate the phase transition from extended eigenstates to localized eigenstates by varying the filling fractions, denoted as $\rho = {n}/N$, in the mobility-edge phase, where $n$ represents the index of the highest occupied eigenstate by fermions.} 

\begin{figure}[t]
\centering
\includegraphics[width=\columnwidth]{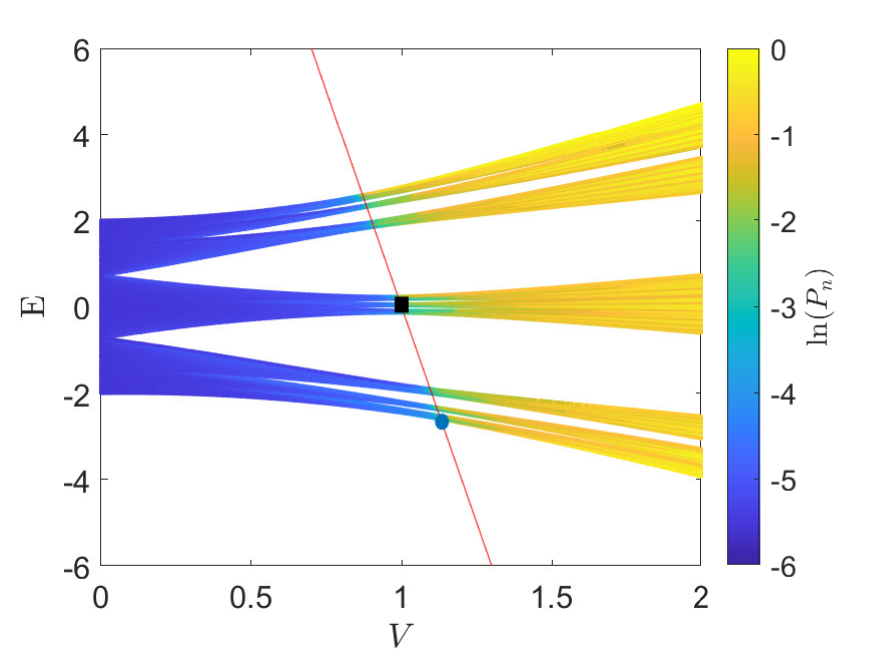}
 \caption{ The eigenvalues of Eq. (\ref{equ:Hamiltonian})
and IPRs as functions of the incommensurate potential strength $V$ with deformation parameter $b=0.1$ and system size $N=1597$.
The eigenvalue curves are color-coded to represent the varying magnitudes of the logarithmic IPR of their corresponding wave functions. 
The red line delineates the self-dual mobility edge as described in Eq. (\ref{equ:mathsolution}), marking the threshold where the energy-dependent localization transition occurs.  Two symbols, black square and the blue circle, highlight the fermion numbers at ${n}=\lfloor N/2 \rceil$ and ${n}=1$, respectively, to illustrate the scaling behaviors at these specific fillings in the following.}
 \label{fig:IPR phase diagram}
\end{figure}
 
\section{Localized transition of GAA model with exact mobility edges}
\label{sec:QPTGAA} 
In order to obtain the eigenenergies of the system
and the associated eigenstates on a finite lattice containing $N$ sites, we numerically diagonalize the model Hamiltonian in Eq. (\ref{equ:Hamiltonian}). We impose antiperiodic boundary conditions  
to alleviate the boundary effects{~\cite{Fraxanet2021,Cookmeyer2020}}. 
In the actual numerical calculations, it is wise to
replace the inverse golden ratio with a rational approximant being $F_{k-1}/F_{k}$, which will approach the inverse golden
ratio for $k \to \infty$, i.e.,  $\alpha=\lim_{k\to \infty} F_{k-1}/F_{k}$.  Here $F_k$ represents the $k$-th Fibonacci number, which is recursively defined by $F_{k+1}=F_{k}+F_{k-1}$, with $F_{0}=F_{1}=1$. The Fibonacci sequence, ${F_k}$, demonstrates a three-cycle parity pattern: odd, odd, even. As a result, the sequence is partitioned into three distinct subsequences,  $\{F_{3l+1}\vert$ $\cdots, 55,233,987, \cdots\}$, $\{F_{3l+2}\vert$ $\cdots, 89,377,1597, \cdots\}$,  $\{F_{3l+3}\vert$ $\cdots, 144,610,2584, \cdots\}$.

At the $k$-th rational approximation, the GAA potential and the Schr\"{o}dinger equation have a periodicity of 
$F_k$ sites, allowing for experimental realizations with available laser wavelengths. 
Then, it is convenient
to calculate the physical quantities of interest on finite lattices. 
The eigenspectra and eigenstates 
can be determined by solving the Schr\"{o}dinger equation 
\begin{eqnarray}
\hat{H} |\Psi_n\rangle=E_{n}|\Psi_n\rangle,
\end{eqnarray}
where $\ket{\Psi_n}=\sum_j \psi_{n,j}\vert j \rangle $ expressed in terms of Wannier basis $\vert j \rangle$, representing the state of a single particle localized at the site
$j$ of the lattice. The Hamiltonian (\ref{equ:Hamiltonian}) can be recast into \begin{eqnarray}
\hat{H}= \hat{\mathbf{c}}^\dagger \mathcal{H} \hat{\mathbf{c}},
\end{eqnarray} 
where $\hat{\mathbf{c}}=[\hat{c}_{1},\cdots,\hat{c}_{F_k}]^{T}$ and $\mathcal{H}$ is a $F_k \times F_k$ single-particle Hamiltonian matrix, given by
\begin{eqnarray}
\mathcal{H}=\left(
\begin{array}{ccccccc}
V_1&-t&0&\cdots&\cdots&\cdots&t\\
-t&V_2&-t&0&\cdots&\cdots&0\\
0&-t&V_3&-t&\ddots&\ddots&0\\
\vdots&\ddots&\ddots&\ddots&\ddots&\ddots&\vdots\\
0&\cdots&0&-t&V_{F_k-2}&-t&0\\
0&\cdots&\cdots&0&-t&V_{F_k-1}&-t\\
t&\cdots&\cdots&\cdots&\cdots&-t&V_{F_k}
\end{array}
\right).
\end{eqnarray}
In order to visually characterize the localized and extended nature of the entire energy spectrum, we evaluate the normalized inverse participation ratio  (IPR) for $n$-th eigenstate $\ket{\Psi_{n}}$ of the model (\ref{equ:Hamiltonian}), given by~\cite{Misguich2016,Licciardello1978,Evers2008}
\begin{eqnarray}
P_n=\frac{\sum_{j=1}^N|\psi_{n,j}|^4}{(\sum_{j=1}^N|\psi_{n,j}^2|)^2}.
\label{equ:IPR} 
\end{eqnarray}
The IPRs can quantify the extent of distribution over the Fock space. For a set of single-particle states in real space, the IPR scales inversely with the system size $N$ in extended states, while appears to be independent of $N$ in the localized states.
We plot the logarithm of normalized IPR for all eigenmodes of Eq. (\ref{equ:Hamiltonian}) as functions of $V$ with the deformation parameter $b=0.1$ in Fig. \ref{fig:IPR phase diagram}. {Hereafter, unless otherwise specified, we set $b=0.1$.} The localization transition from the EP to the LP occurs at mobility edges in Eq. (\ref{equ:mathsolution}). In the following, our primary emphasis will be on two specific cases: one where the fermion number is ${n}=\lfloor N/2 \rceil$, corresponding to a filling fraction of $\rho=1/2$, and another where the fermion number is ${n}=1$, corresponding to a filling fraction of $\rho=1/N$.  Here $\lfloor \cdot\rceil$  denotes the round function.

\begin{figure}[t]
\centering
\includegraphics[width=\columnwidth]{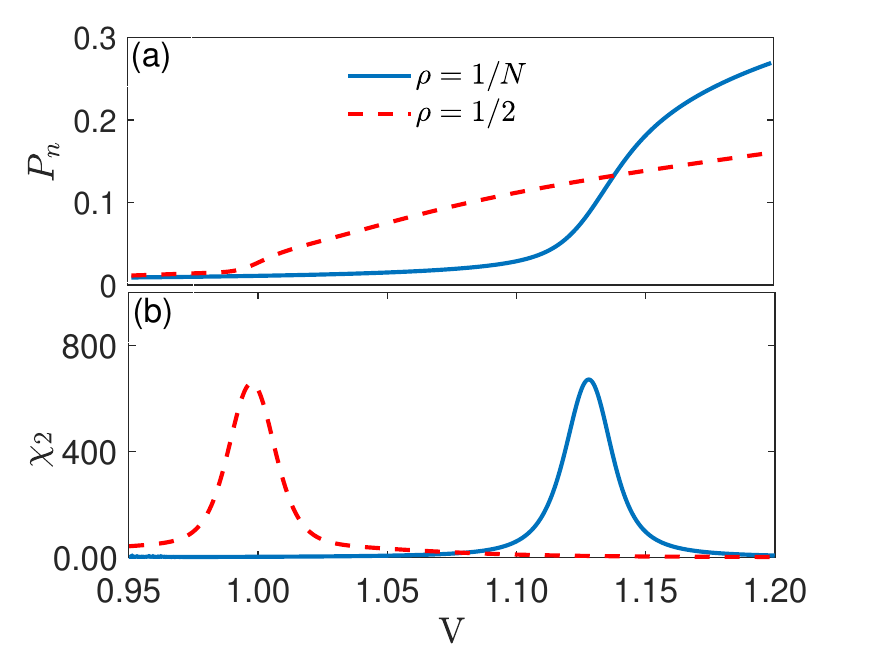}
 \caption{(a) IPRs and (b) QFS as a function of the incommensurate potential strength $V$ for different filling fractions: $\rho=1/2$ (red dashed lines) and  $\rho=1/N$ (blue dashed lines). Here we choose the deformation parameter $b=0.1$,  phase $\phi=\pi$  and the lattice site $N=377$. }
\label{fig:IPR_FS_b_0.1}
\end{figure}

As the IPR is bounded, it is not easy to accurately identify the critical point, as is revealed in Fig. \ref{fig:IPR_FS_b_0.1}(a).
Here we use quantum fidelity susceptibility (QFS)  as a versatile indicator of localization transition points. 
The fidelity susceptibility quantifies the sensitivity of an eigenstate to variations of the driving parameter, which is defined as the second-order derivative of the quantum fidelity with respect to the driving parameter~\cite{You2007}.  Being a purely geometric measure, the fidelity approach offers a notable advantage by not requiring any prior knowledge of the order parameter. 
A GFS of order $2r + 2$ associated with the eigenstate $\ket{\Psi_{n}}$ with the tuning parameter $V$ is given by~\cite{You2015}
\begin{eqnarray}
\chi_{2r+2}^{(n)}=\sum_{m\neq n}\frac{|\bra{\Psi_m}\partial_{V}\hat{H}\ket{\Psi_n}|^2}{(E_m-E_n)^{2r+2}}.
\label{generalizedsusceptibility}
\end{eqnarray}
The numerator in Eq.(\ref{generalizedsusceptibility}) denotes the probability of exciting the system away
from the state $\vert \Psi_n \rangle$ through a relevant (or marginal) perturbation $\partial_V \hat{H}$. The GFS of different orders is encoded  by  the  power of the denominator.  
For the case of $r=-1/2$, Eq. (\ref{generalizedsusceptibility}) simplifies to the second derivative of the eigenenergy~\cite{Chen2008}, denoted as $\chi_1^{(n)} \equiv \partial^2 E_n/\partial V^2$. Similarly, when $r=0$, Eq. (\ref{generalizedsusceptibility}) reverts to the QFS, i.e., $\chi_2^{(n)} \equiv \partial^2 (1-F)/\partial V^2$, where the fidelity $F$ is defined as the absolute value of the inner product between $\vert \Psi_n (V) \rangle$ and $\vert \Psi_n (V + \delta V) \rangle$. This fidelity quantifies the Bures distance between two states as their tuning parameter slightly changes.
In the following, we focus on the GFS of the eigenstate $\vert\Psi_n (V)\rangle $ at a finite filling fraction for $n=\lfloor\rho F_k\rceil$, where $\lfloor x\rceil$ rounds $x$ to the nearest integer.
To this end, the superscript of  $\chi_{2r+2}^{(n)}$ is omitted for abbreviation.

Different from the smooth transition of the IPR around the mobility edge, the QFS shows a divergent peak in the vicinity of the EP-LP transition point due to 
the enhanced sensitivity, as is disclosed in Fig. \ref{fig:IPR_FS_b_0.1}(b).  The QFS and GFS have been widely used to witness the quantum phase transitions in quantum systems~\cite{GU2010FIDELITY} and even holographic models~\cite{Momeni2017,LeBlond2021}.  
They have proved to be particularly useful for detecting the critical points of a symmetry-knowledge unknown systems~\cite{Wei2018,Garnerone2009,Mao2021}. 
It has been demonstrated that the QFS not only facilitates the identification of quantum critical points (QCPs) but also adheres to the scaling ansatz, enabling the retrieval of critical exponents. The retrieved critical exponents conform to  the hyperscaling relations, signifying the existence of only two independent exponents. The key implication is that the finite-size scaling behavior of the standard fidelity susceptibility $\chi_2$ determines the positions of  QCP, along with the correlation-length critical exponent $\nu$ associated. Notably, a second critical exponent assumes a pivotal role in establishing the universality class such as the dynamical exponent $z$. In the context of quasiperiodic systems exhibiting spatial complexity, the scaling behavior of the QFS  and the universality class of the localization transition have received relatively less exploration.
Specifically, we are able to extract two critical exponents and universal information by observing the finite-size scaling of the GFS, which is particularly remarkable considering the constraints posed by the limited number of system sizes available within quasiperiodic systems.

\begin{figure*}[t]
\centering
\includegraphics[width=0.8\columnwidth]{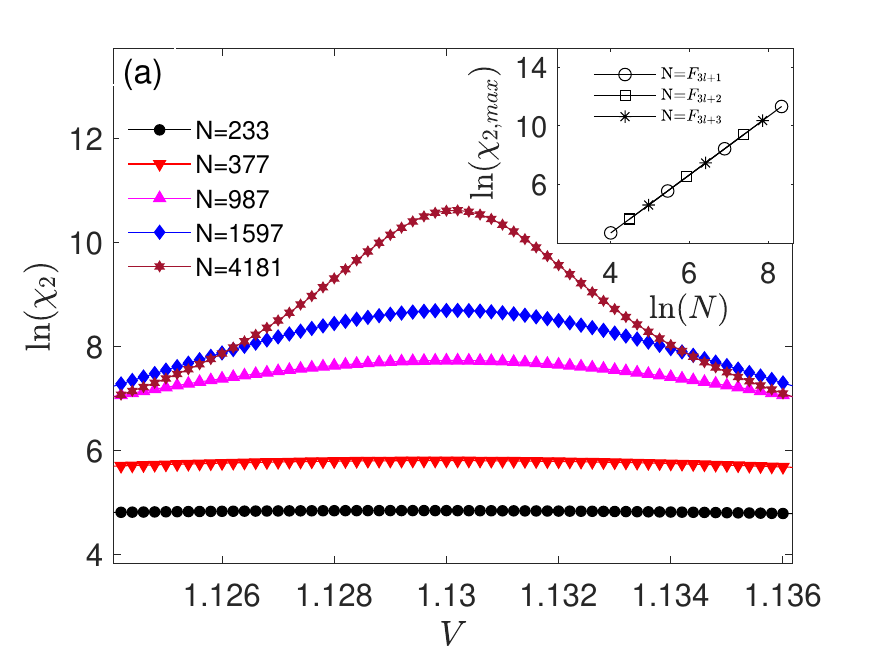}
\includegraphics[width=0.8\columnwidth]{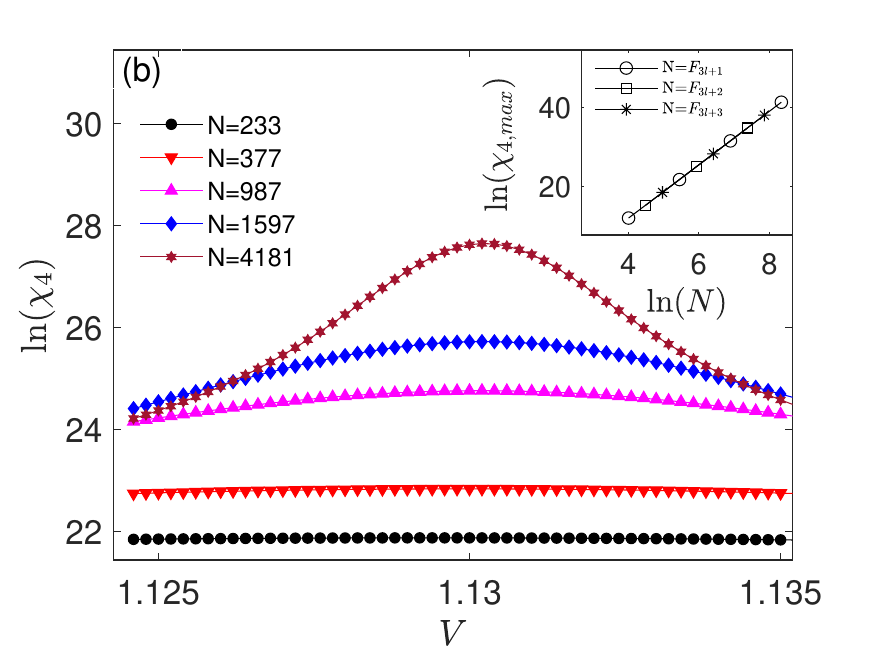}
\includegraphics[width=0.8\columnwidth]{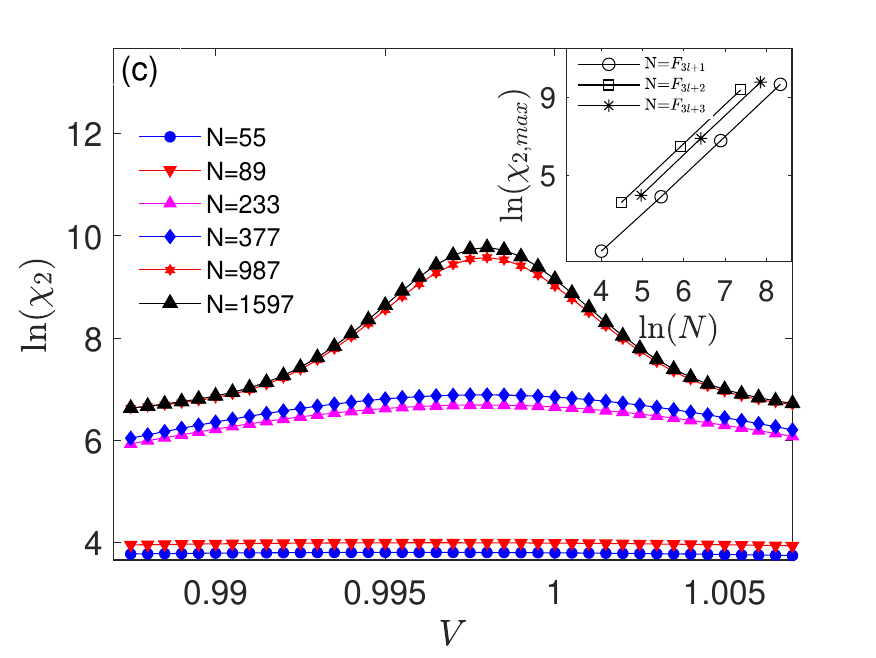}
\includegraphics[width=0.8\columnwidth]{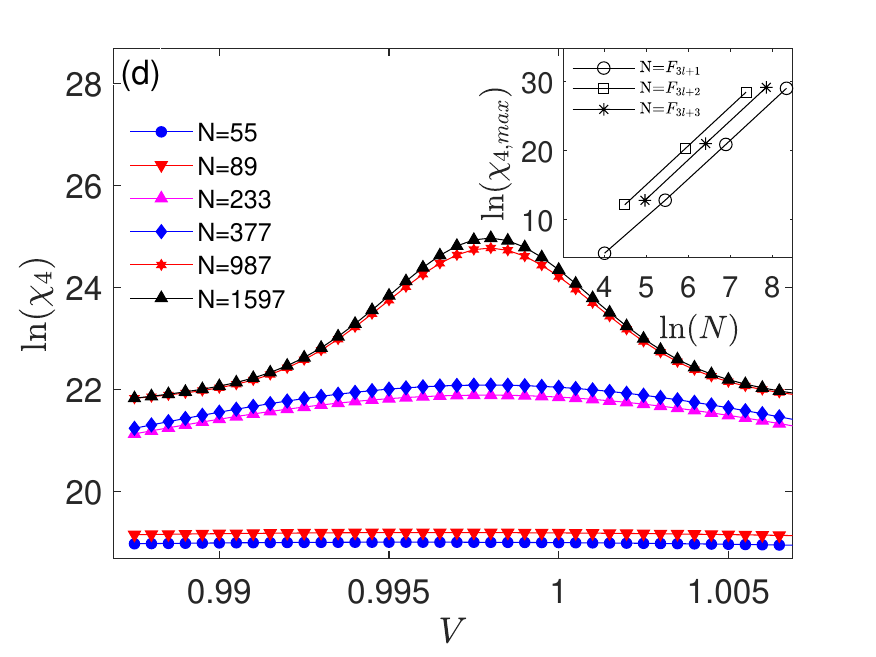}
\caption{The GFS in the GAA model with deformation parameter $b=0.1$ and phase $\phi=\pi$. (a) The logarithm of the second-order GFS $\chi_{2}$ and (b) the fourth-order GFS $\chi_4$ as functions of the incommensurate potential strength $V$ for various lattice sizes at a filling fraction of $\rho=1/N$. Panels (c) and (d) similarly depict $\chi_{2}$ and $\chi_4$, respectively, at a filling fraction of $\rho=1/2$. The insets in each panel illustrate the scaling behavior of the peak maxima for system sizes represented by Fibonacci sequences $N=$ $\{F_{3l+1}\vert 55, 233, 987\}$, $\{F_{3l+2}\vert 89, 377, 1597\}$ and $\{F_{3l+3}\vert 144, 610, 2584\}$.}
\label{fig:FS_scaling_rho_b_0.1}
\end{figure*}

\section{Scaling behavior and Diophantine equation}
\label{sec:Scaling hypothesis}
 
The finite-size scaling theory of GFS  is applicable in the vicinity of continuous quantum phase transitions, as the sensitivity is much more pronounced for the GFS around the QCP than that away from criticality~\cite{Braun2018}. 
{When $V$ is adiabatically crossed over the QCPs $V_c$, the GFS of a finite system displays a peak, effectively indicating the positions of pseudocritical points $V_m$. 
With increasing system sizes $N$, the peaks of the GFS become
more prominent.} Additionally, it is anticipated that the maximal value of the GFS is expected to obey a power-law relation with the system size as~\cite{You2015}
\begin{eqnarray}
\chi_{2r+2}(V_m) \sim N^{-\mu},
\label{equ:scaling1}
\end{eqnarray}
where $\mu=2/\nu+2zr$ corresponds to the critical adiabatic dimension. {Here, $\nu$ denotes the correlation-length exponent, and $z$ represents the dynamical critical exponent. According to single-parameter scaling theory, the correlation length $\xi$ becomes the only relevant length scale in the thermodynamic limit that diverges near the phase transition as~\cite{Fisher1972},
\begin{eqnarray}
\xi\sim|V-V_c|^{-\nu},
\end{eqnarray}
 and the single-particle spectral gap $\epsilon_s$ is predicted to close following the relation
\begin{eqnarray}
\epsilon_s\sim N^{-z}.
\end{eqnarray} 
For} a relevant operator such as $\partial_{V}\hat{H}$ applied to sufficiently large one-dimensional systems, where $\nu \textless 2$, the pseudocritical points $V_m$ tend to approach the critical points $V_c$, thereby satisfying 
\begin{eqnarray}
|V_m-V_c| \propto N^{-1/\nu}.
\label{equ:scaling2}
\end{eqnarray}
Consequently, the GFS of a finite system with size $N$ in the vicinity of a QCP shall obey a universal scaling form ~\cite{Albuquerque2010},  
\begin{equation}
\chi_{2r+2}(V)=N^{2/\nu+2zr}\phi_r(|V-V_m|N^{1/\nu}).
\label{equ:scaling3}
\end{equation}
Here, $\phi_r (x)$ represents a unknown regular universal scaling function for the GFS of order $2r+2$. In this case, estimates for critical parameters can be deduced by plotting the scaled GFS as $[\chi_{2r+2}(V_m)-\chi_{2r+2}(V)]/\chi_{2r+2}$ against $N^{1/\nu}(V-V_{m})$, with subtle adjustments made to the values of $V_m$, $\nu$, and $z$ until achieving a collapse of data.

\par
Figure \ref{fig:FS_scaling_rho_b_0.1} shows the logarithm of the GFS, including $\ln \chi_2$ and $\ln\chi_4$ of the GAA model, as a function of $V$ for various system sizes $N$ with $\phi=\pi$. 
We observe that $\chi_4$ exhibits a much more pronounced divergence compared to $\chi_2$ at a critical point, yet there exist distinctions between the cases of $\rho=1/N$ and $\rho=1/2$.
The logarithmically scaled maximum values of the GFS, $\chi_{\rm 2,max}$ and $\chi_{\rm 4,max}$, in proximity to the QCP are plotted against various system sizes $N$, which confirm the linear scaling relations by the numerical fittings. In-depth analysis of the insets in Figs. \ref{fig:FS_scaling_rho_b_0.1}(a) and (b) reveals interesting behavior regarding the filling fraction $\rho=1/N$. Specifically, at this fraction, only a single subsequence is apparent. However, an interesting deviation occurs for even subsequences compared to odd ones when the phase $\phi$ is set to 0. To mitigate this odd-even effect and ensure consistent data analysis, we appropriately adjust the value of  $\phi$ to $\pi$, thereby aligning the subsequences more uniformly.
 According to Eq. (\ref{equ:scaling3}), the linear fitting for various system
sizes $N$ gives rise to $\nu=0.998\pm0.0015$ and $z=2.3705\pm0.0025$, while the numerical fitting for even number of system sizes $N$ yields $\nu=0.998\pm0.009$ and $z=2.371\pm0.018$.
On the contrary, when considering $\rho=1/2$, the numerical result reveals a division of the sequence into three distinct subsequences for numerous values of  $\phi$. 
As shown in the inset of Fig. \ref{fig:FS_scaling_rho_b_0.1}(c), 
the linear fittings of the log-log plot give rise to $\nu=1.004\pm0.0515$ for $N=F_{3l+1}$, $\nu=0.9985\pm0.0030$ for $N=F_{3l+2}$ and $\nu=1.0005\pm0.0015$ for $N=F_{3l+3}$. Likewise, from the inset of Fig. \ref{fig:FS_scaling_rho_b_0.1}(d), the  dynamical exponent can be fitted as $z=1.8265\pm0.0082$ for $N=F_{3l+1}$, $z=1.8305\pm0.0138$ for $N=F_{3l+2}$ and $z=1.8355\pm0.078$ for $N=F_{3l+3}$. 
By calculating their averages, we determine the correlation-length critical exponent of $\nu=1.001\pm0.0186$ and the dynamical critical exponent of $z=1.83083\pm0.03332$ for the filling fraction $\rho=1/2$. 

{Recent studies have highlighted an intimate relation between finite-size effects and Diophantine equation in the AA model~\cite{Cookmeyer2020}.  The solutions of the Diophantine equation, obtained through continued fraction expansions, elucidate a mathematical framework that captures the essence of the system's quasi-periodicity, which is fundamental to understanding its renormalization group dynamics~\cite{Szabo2018}.} 
The Diophantine equation establishes a link between the filling fraction and the disorder's incommensurate frequency, refining the resonance criterion $Q \alpha-P = \rho$ of commensurate cases~\cite{PhysRevB.65.115114},
where $Q$ and $P$ are integers that do not exceed the system size $N$. This association is based on the continued-fraction expansion of the frequency $\alpha$.  When $Q$ and $P$ remain fixed irrespective of system size, and the resonance equation is solved precisely, the system is deemed commensurate. In this state, $Q$ determines the perturbation order where the resonance between the Fermi momentum and the potential frequency $\alpha$ occurs. Conversely, when $Q$ and $P$ scale with the system size, the system is recognized as incommensurate. Under these circumstances, the resonance criterion transcends into a Diophantine equation requiring a solution that accommodates the scaling variables, which then forms the mathematical foundation for understanding 
the resonance phenomena within the incommensurate framework. The incommensurate filling $\rho$, marked by the growth of $Q$ and $P$ with system size $N$, implies a dynamic interplay between the system's physical dimensions and its spectral properties. This Diophantine equation not only serves as a tool for solving for these properties but also encapsulates the essence of the incommensurate state within the GAA model. Our investigation focuses on exploring this relationship further within the GAA model, particularly examining the sequence effects. 

We note that $\alpha$  can be approximated by $\alpha_k=F_{k-1}/F_{k}$ and ${n}=\lfloor \rho F_{k}\rceil$. Consequently, the Diophantine equation is reformulated as 
\begin{eqnarray}
Q_k F_{k-1}-P_k F_{k}= \pm {n},
\label{equ:Diophantine equation}
\end{eqnarray}
To seek integer solutions ($Q_k$, $P_k$) for given integer values of $F_k$ and ${n}$. For each value of $k$, an infinite spectrum of solutions ($Q_k$, $P_k$) hat satisfy the resonance condition  in  Eq. (\ref{equ:Diophantine equation}). 
The solutions are constrained to  $|Q_k|\leq F_k/2$,and among these, the smallest $|Q_k|$ is selected. 
 When solving the Diophantine equation for different values of $F_k$ and {$n$}, the ratio $|Q_k|/F_k$ for large $k$ displays a periodicity of $p$ distinct values, thus dividing the sequence of denominators into $p$ subsequences.  As illustrated in Table \ref{table p}, $p$ also corresponds to the number of scaling functions, such as $p=1$ for $\rho=1/N$ and $p=3$ for $\rho=1/2$ at half filling. Further details are available in Appendix \ref{Append1}.

\begin{figure}[ht]
\centering
\includegraphics[width=0.95\columnwidth]{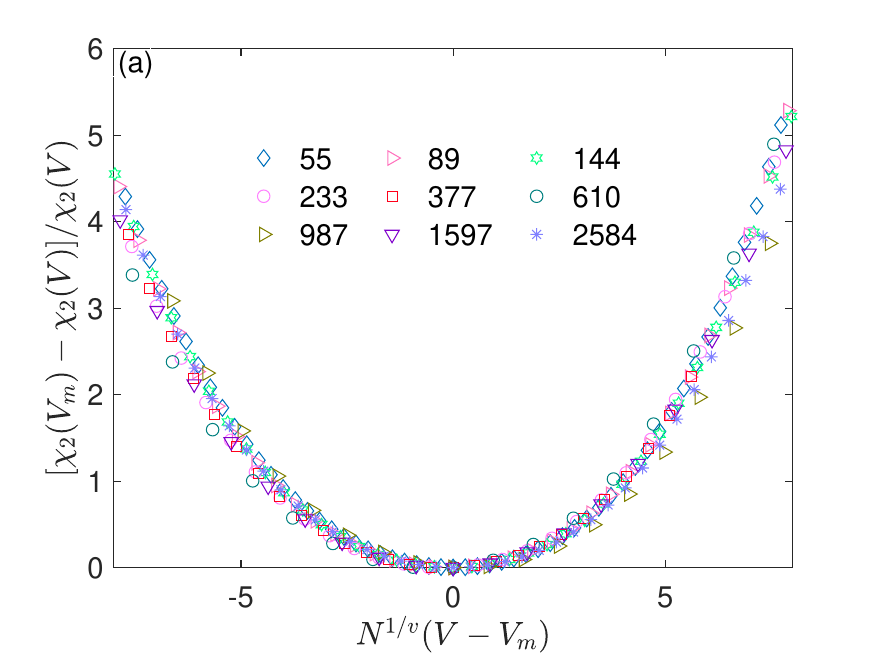}\\
\includegraphics[width=0.95\columnwidth]{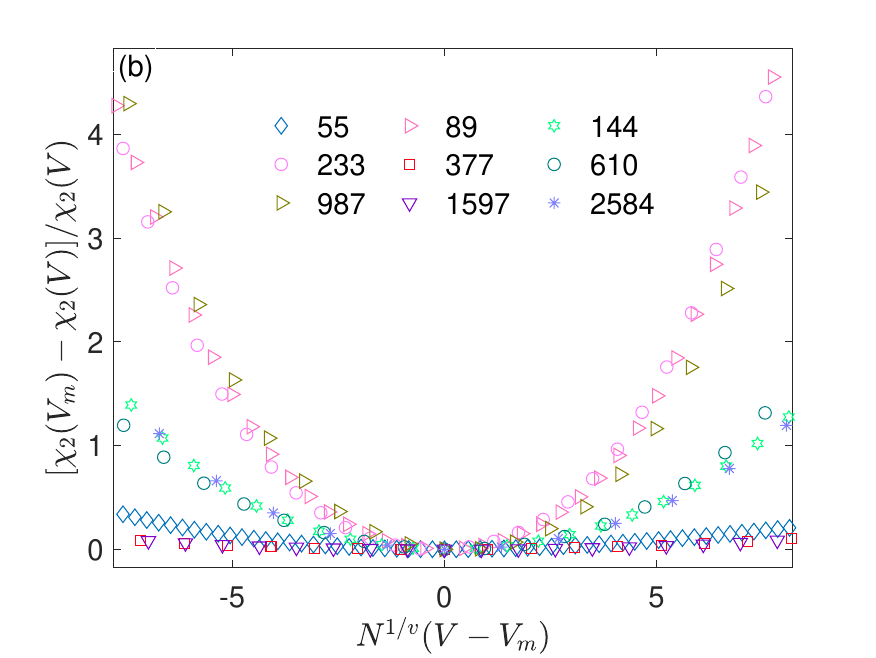}
 \caption{Scaling analysis of the GFS in the GAA model. (a) Scaled fidelity susceptibility, calculated as $[\chi_{2r+2}(V_m)-\chi_{2r+2}(V)]/\chi_{2r+2}$, is plotted against the scaled variable  $N^{1/\nu}(V-V_{m})$ for a filling fraction of $\rho=1/N$.   
(b) A similar analysis is presented for a filling fraction of $\rho=1/2$, where the scaled fidelity susceptibility is plotted against the same scaled variable. Here we choose deformation parameter $b=0.1$, phase $\phi=\pi$ and the correlation-length critical exponent $\nu=1.000$.}
\label{fig:bdhs_FS_b_0.1}
\end{figure}
 
\begin{table}[htbp]
\begin{tabular}{ |c|c|c|c|c|}
\hline
 \diagbox[dir=NW]{$\rho$}{$b$}&0&0.25&0.5&$p$\\
\hline
$1/N$&2.3730$\pm$0.0010&2.3740$\pm$0.0020&2.3755$\pm$0.0040&1\\
\hline
$1/3$&2.0017$\pm$0.0080&2.0046$\pm$0.0090&2.0122$\pm$0.0180&4\\
\hline
$1/2$&1.8285$\pm$0.0075&1.8250$\pm$0.0030&1.8245$\pm$0.0035&3\\
\hline
$1$&2.3740$\pm$0.0020&2.3750$\pm$0.0012&2.3735$\pm$0.0010&1\\
\hline
\end{tabular}
\caption{The dynamical exponents $z$ and numbers of sequences $p$ extracted from finite-size scaling at various fillings. The number of scaling functions $p$ is determined by scaling collapse as in Fig. \ref{fig:bdhs_FS_b_0.1} and via the Diophantine equation where it is determined by the period of the repeating values of $|Q_k|/F_k$ for large $k$.}
\label{table p}
\end{table}

The universality class, central to the study of critical phenomena, posits that associated critical exponents and scaling functions are universal, subject only to symmetries and space dimensionality. We validate the obtained critical exponents using the data collapse technique, as demonstrated in Fig. \ref{fig:bdhs_FS_b_0.1}. The scaling hypothesis suggests that appropriately scaled curves can converge into a single unified curve. To facilitate this data collapse, we employ the rescaled fidelity susceptibility as defined in Eq.(\ref{equ:scaling3}): 
\begin{eqnarray}
\left[\chi_2 (V_m)-\chi_2 (V)\right]/\chi_2 (V)\sim \mathcal{F}_q ( N^{1/\nu}\delta),
\label{equ:scaling4}
\end{eqnarray}
where $\delta =V-V_m$ is the deviation from the pseudocritical point, and $\mathcal{F}_q$ ($q=1,2,\cdots p$) is  scaling functions.  Based on Eq. (\ref{equ:scaling4}),  we plot  $\left[\chi_2 (V_m)-\chi_2 (V)\right]/\chi_2 (V)$ versus $N^{1/\nu}(V-V_m)$ for different values of $N$ as depicted in Fig.\ref{fig:bdhs_FS_b_0.1}. For  $\rho=1/N$  all curves of distinct chain sizes in the vicinity of $V_m$ collapse onto a single curve in Fig. \ref{fig:bdhs_FS_b_0.1} (a), while  for $\rho=1/2$ the scaling of $\chi_2$ collapses all the curves onto three scaling functions.  
 A very good quality data collapse confirms that the scaling hypothesis is in place and the estimated value of $V_c$, $\nu$, $z$ obtained through the prescriptions in Eq.(\ref{equ:scaling1}) and (\ref{equ:scaling2}) are self-consistent.

Similar strategies  are applicable across various filling fractions. Our main findings in this work are the numerical results of the dynamical exponent $z$ for different filling $\rho$ and  various values of  $b$, as summarized in Tab. \ref{table p}.  We observe that the value of $z$ and the number of scaling functions $p$  are influenced by the filling fraction $\rho$, yet they remain independent of the deformation parameter $b$.  
As indicated in Tab.\ref{table p},  for a filling fraction of $\rho=1/N$, the dynamical critical exponent $z$ is approximately 2.734. At $\rho=1/3$, $z$ is around 2.00, and for $\rho=1/2$, $z$ is approximately 1.825. Intriguingly, at $\rho=1$, despite the localization transition not occurring at the same critical points as in the case of $\rho=1/N$,  a symmetry persists, and the dynamical critical exponent is approximately 2.734 in both scenarios. This consistency of the dynamical critical exponents across similar filling states under varying mobility edges demonstrates that the Diophantine equation is applicable to the AA model, even when self-duality is broken.

\section{Summary}
\label{Summary and conclusions}
In this study, we employ quantum fidelity susceptibility (QFS) to pinpoint the mobility edge in the Anderson localization transition. Recognized as a potent tool for probing quantum criticalities, the fidelity susceptibility has demonstrated its efficacy in one-dimensional disordered lattice models, as well as in the p-wave-paired Aubry-Andr\'{e}-Harper model. Here our research thus focuses on the generalized Aubry-Andr\'{e} (GAA) model, which is characterized by an exact mobility edge (ME) leading to a transition between the localized phase and the extended phase within its energy spectrum. A significant aspect of our approach involves utilizing generalized fidelity susceptibility (GFS) to deduce both the correlation-length critical exponent $\nu$ and the dynamical critical exponent $z$. A notable observation in our study is the manifestation of finite-size effects on the critical behavior of GFS. Furthermore, the number of scaling functions, denoted as $p$, conforms to the Diophantine equation, taking into account the varying filling fractions and system sizes. This led us to conduct extensive tests across different filling fractions and under diverse ME conditions, culminating in the derivation of the correlation-length and dynamical critical exponents, as shown in Table \ref{table p}. The recent experimental verification of the existence of MEs~\cite{Wang2022} underscores the timeliness and relevance of our theoretical approach. By elucidating the localization transformations and critical exponents in the GAA model via QFS, our work lays a robust theoretical foundation for understanding the underlying mechanisms of ME phenomena. 
In conclusion, this research makes a substantial contribution to the field of quantum physics, particularly in understanding phenomena like the localization transition, mobility edge, and critical behavior in quantum systems. It introduces novel analytical methodologies and provides theoretical insights that align closely with recent experimental findings, thus enriching the current scientific discourse in this domain.

 \begin{acknowledgments}
The authors appreciate very insightful discussions with Ting Lv. This work is supported by the National Natural Science Foundation of
China (NSFC) under Grant No. 12174194, Postgraduate Research \&
Practice Innovation Program of Jiangsu Province, under Grant No. KYCX23\_0347, Opening Fund of the Key Laboratory of Aerospace Information Materials and Physics (Nanjing University of Aeronautics and Astronautics), MIIT, Top-notch Academic Programs Project of Jiangsu Higher Education Institutions (TAPP), and stable supports for basic institute research under Grant No. 190101. 
\end{acknowledgments}

\appendix
\section{Number of scaling functions determined by the Diophantine equation}
\label{Append1}
 
The universality class of the single-particle ground state is known to depend solely on the continued-fraction expansion of the wave vector $\alpha$. This expansion can be expressed as 
\begin{equation}
\alpha = {\alpha}_1 + \frac{1}{{\alpha}_2 + \frac{1}{{\alpha}_3 + \ddots}} = [{\alpha}_1, {\alpha}_2, {\alpha}_3, \cdots].
\end{equation}
By truncating the series at ${\alpha}_k$ we obtain a rational approximation of $\alpha$ as $\alpha \approx M_k/N_k$,  where $M_k$ and $N_k$ are coprime. Consequently the system sizes are considered to be  $N_k$ that satisfy (anti)periodic boundary conditions. Generally, two values of  $\alpha$ are said to have the same asymptotic continued-fraction expansion if there exists a natural number $k$ such that for all $i>k$, the $n_i$ in their expansions are identical.
The number of scaling functions $p$ is determined by the scaling collapse of the GFS and also by the Diophantine equation. This is particularly influenced by the period of the repeating values of $q_k\equiv |Q_k|/N_k$ for large $k$, for specific values of the wave number $\alpha$ and the filling fraction $\rho$. For instance, the golden mean $\alpha=(\sqrt{5}+1)/2$=$[1,1,1,\cdots]$. The inverse of the golden mean is $\alpha = (\sqrt{5}-1)/2$=$[0,1,1,\cdots]$. Its rational approximations from the continued-fraction expansion are given by $\alpha \approx  F_{k+1}/F_{k}$, 
where $F_{k}$ are the Fibonacci numbers, defined by $F_1$ = 1, $F_2 = 1$, and $F_{k+1} = F_k +  F_{k-1}$.

Considering a filling fraction the filling fraction $\rho=1/N$ (i.e., ${n}=1$), we can calculate $Q_k$ and $P_k$ for different system sizes $F_k$. The following equations present these calculations:
\begin{equation}
\begin{aligned}
&F_8=21,{n}=1,Q_8=8,P_8=5,q_8=0.38095,\\ \nonumber
&F_9=34,{n}=1,Q_9=13,P_9=8,q_9=0.38235,\\ \nonumber
&F_{10}=55,{n}=1,Q_{10}=21,P_{10}=13,q_{1}=0.38182,\\ \nonumber
&F_{11}=89,{n}=1,Q_{11}=34,P_{11}=21,q_{1}=0.38202,\\ \nonumber
&F_{12}=144,{n}=1,Q_{12}=55,P_{12}=34,q_{1}=0.38194,\\ \nonumber
&F_{13}=233,{n}=1,Q_{13}=89,P_{13}=55,q_{1}=0.38197,\\ \nonumber
&F_{14}=377,{n}=1,Q_{14}=144,P_{14}=89,q_{1}=0.38196, \\\nonumber 
&F_{15}=610,{n}=1,Q_{15}=233,P_{15}=144,q_{1}=0.38197,\\ \nonumber
&F_{16}=987,{n}=1,Q_{16}=377,P_{16}=233,q_{1}=0.38197 ,\\ \nonumber
&F_{17}=1597,{n}=1,Q_{17}=610,P_{17}=377,q_{1}=0.38197. \nonumber
\end{aligned}
\end{equation}
It becomes apparent that the value of $|Q_k|/N_k$ for large $k$ stabilizes with a period of $p=1$ in the case of $\rho=1/N$.

Now, considering the half filling, i.e., $\rho=1/2$, we again determine $Q_k$ and $P_k$ for various system sizes $F_k$. The data is presented as follows:
\begin{equation}
\begin{aligned}
&F_8=21,{n}=11,Q_8=4,P_8=3,q_8=0.19048,\\ \nonumber
&F_9=34,{n}=17,Q_9=17,P_9=10,q_9=0.5,\\ \nonumber
&F_{10}=55,{n}=28,Q_{10}=17,P_{10}=10,q_{2}=0.30909,\\ \nonumber
&F_{11}=89,{n}=45,Q_{11}=17,P_{11}=10,q_{3}=0.19101,\\ \nonumber
&F_{12}=144,{n}=72,Q_{12}=72,P_{12}=44,q_{4}=0.5,\\ \nonumber
&F_{13}=233,{n}=117,Q_{13}=72,P_{13}=45,q_{2}=0.30901,\\ \nonumber
&F_{14}=377,{n}=189,Q_{14}=72,P_{14}=45,q_{3}=0.19098,\\ \nonumber
&F_{15}=610,{n}=305,Q_{15}=305,P_{15}=188,q_{4}=0.5,\\ \nonumber
&F_{16}=987,{n}=494,Q_{16}=305,P_{16}=188,q_{2}=0.30902,\\ \nonumber
&F_{17}=1597,{n}=799,Q_{17}=305,P_{17}=188,q_{3}=0.19098. \nonumber
\end{aligned}
\end{equation}
In this case, the value of $|Q_k|/N_k$  for large $k$ exhibits a periodicity of $p=3$, indicating different values and breaking the sequence of denominators into three subsequences in the case of $\rho=1/2$.

\bibliography{ref}
\end{document}